\documentclass[12pt]{article}
\usepackage{amsmath,amssymb}
\def\R{\hbox{{\rm I}\kern-0.2em{\rm R}\kern0.2em}}%definition of reals
\def\R{\hbox{{\rm I}\kern-0.2em{\rm R}\kern0.2em}}%mathematical R for reals
\def\D{\hbox{{\rm I}\kern-0.2em{\rm D}\kern0.2em}}

\def\be{\begin{equation}}
\def\ee{\end{equation}}

\def\({\left(}
\def\){\right)}
\def\[{\begin{equation}}
\def\]{\end{equation}}
\def\bc{\begin{center}}
\def\ec{\end{center}}

\usepackage{graphicx}
\usepackage{epstopdf}
\usepackage{color}

\begin{document}

%{\large \bf Marginally Stable Circular Orbits of a Test Particle in Schwarzschild Black Hole Geometry Surrounded by Quintessence Matter}
{\large \bf Marginally Stable Circular Orbits in Schwarzschild Black Hole Surrounded by Quintessence Matter}
%{\large \bf The Impact of Quintessence on the Marginally Stable Circular Orbits of a Test Particle in Schwarzschild Black Hole}
%{\large \bf On the Stability of Orbits of a Test Particle in Schwarzschild Black Hole Surrounded by Quintessence}

\textit{Ibrar Hussain, Sajid Ali}

Dept. of Basic Sciences, \\
School of Electrical Engineering and Computer Science,\\
National University of Sciences and Technology,
H-12 Campus, Islamabad, Pakistan\\

E-mail: ibrar.hussain@seecs.nust.edu.pk; sajid$\_$ali@mail.com

{\bf Abstract}. Marginally stable circular orbits (MSCOs) of a massive test particle
are investigated in the spacetime geometry of Schwarzschild black hole
surrounded by quintessence. For that matter we consider three important scenarios where the equation of
state parameter $\omega_{q}$, has one of the following forms (i) $\omega_q=-1$ (ii) $\omega_q=-2/3$ and (iii) $\omega_q= -1/3$. The existence of such marginally stable circular orbits in these scenarios depend on the range of normalization factor $\alpha$. Briefly, we show that in the first case such orbits exist only if $0<\alpha<4/16875$. Moreover in the second case which is a special Kiselev black hole it is found that MSCOs exist when the value of the normalization factor satisfy $0<\alpha\leq 0.00536165238$. In the last case the MSCOs are also shown to exist.

\textit{Key words}: Marginally stable circular orbits;
Schwarzschild black hole; Quintessence matter; Normalization factor.

\section{Introduction}

Motion of particles in the spacetime geometry of black holes is an
active topic of research for theoretical physicists. It may be helpful
in understanding gravitational field around black holes. In this regard
several black hole spacetimes have been investigated in the literature
\cite{SH}-\cite{	TTN}. In the theory of general relativity the radius of circular
orbits of particles in the vicinity of black holes has a lower bond which
are known as inner most stable circular orbits (ISCOs). While the circular
orbits with the upper bound on its radius are called outer most stable circular
orbits (OSCOs). These two types of orbits form a boundary between the two regions, i.e., a stable region and an unstable region respectively. In the literature this boundary is known as the MSCO \cite{TTN}.
If in a spacetime geometry the number of MSCOs is only two then the smaller is
called the ISCO and the larger is known as the OSCO. On the other hand
if the number of MSCOs is greater than two then the smallest will be the
ISCO and the largest will be the OSCO. A study of the ISCO for the Schwarzschild
black hole can be found in the literature (see fre example \cite{MTW,LL}).

The existence of ISCOs may play an important role in the study of gravitational waves \cite{TTN2}.
It is believed that a super massive black hole exists at the centre of each galaxy
\cite{SMB1,SMB2, SMB3}. In the process of generation of gravitational waves, ISCOs are considered to
be the location where an orbiting compact object e.g, another black hole orbiting
around the super massive black hole goes from the inspiralling phase to the merging
phase \cite{TTN2,SMB4}. ISCOs have their own importance in high energy astrophysics where
the existence of these orbits is related to the inner edge of the accretion
disks around black holes \cite{TTN3}. Therefore, the study of the ISCOs can
provide useful information about the nonlinear spacetime geometry which is
beyond the local tests of our solar system. The OSCO for a test particle  in
the vicinity of Kottler black hole spacetime has been investigated by Stuchlik
and Hledik \cite{SH}.

Cosmological observations like the Supernovae Ia, the Cosmic Microwave Background
radiation anisotropies and X-ray experiments support the accelerated expansion of
our Universe \cite{Pe, PB, DN}. It is believed that dark energy is responsible for
this accelerated expansion of our Universe. Several phenomenological models have been proposed to describe dark energy of which there is one model which examine the possibility of the presence of a scalar field known as quintessence (see e.g., \cite{Mir}).
This scalar field is defined by the equation of negative state parameter,
which is the ratio of the pressure and density \cite{LZ}. Kiselev derived a black
hole solution of the Einstein field equations with quintessence matter
\cite{VVK}. This solution reduces to the Schwarzschild solution of the Einstein
field equations when the quintessence term disappears. Null geodesics for the
Schwarzschild black hole surrounded by quintessence have been investigated by
Sharmanthie Fernando for particular values of the equation of state parameter
$\omega_q=-2/3$ and the normalization factor $\alpha=0.1$, $0.01$, $0.005$
\cite{SF} (the normalization factor is given in the metric coefficient of the time element of the quintessence black hole in the third section). For the same black hole time-like geodesics have been studied by
Rashmi et. al \cite{RNH}. They have considered three different values of the
equation of state parameter $\omega_q=-1$, $-2/3$, $-1/3$ and four different
values of the normalization factor $\alpha=0.1$, $0.08$, $0.05$, $0.005$. For
unit mass of the black hole with $\omega_q=-1/3$, $-2/3$ and $\alpha=0.1$
Rashmi et. al have shown that the radius of the ISCO has shifted to a larger
distance from centre as compared to the pure Schwarzschild black hole.
In the present study we are interested to analyze the MSCOs in the Schwarzschild
back hole surrounded by quintessence scalar field for three different
cases of the equation of state parameter $\omega_q=-1$, $-2/3$, $-1/3$. In
particular we obtain upper and lower bounds on the value of the normalization
factor $\alpha$ to find the radius of the MSCOs in the background of above geometry.

In the next Section we give the necessary and sufficient conditions
for the existence of MSCOs in the case of a spherically symmetric static spacetime.
In Section 3 we study MSCOs in the spacetime geometry of the
Schwarzschild back hole surrounded by quintessence matter. A summary
of the discussion is given in the last Section. In this paper we use
$G=c=1$. From here onwards we refer qSBH as a Schwarzschild black hole surrounded by quintessence.

\section{Necessary and Sufficient Conditions for MSCOs}

The line element for the general spherically symmetric static spacetime is given by
\cite{MTW}
\begin{equation}\label{smetric}
 \mbox{d}s^2=-f(r)\mbox{d}t^2+\frac{1}{f(r)}\mbox{d}r^2+h(r)(\mbox{d}{\theta}^2+\sin^2{\theta}\mbox{d}\phi^2 )\,,
\end{equation}
where $f(r)>0$ and $h(r)>0$, resulting in the signature (-,+,+,+) of spacetime. In the above line element, $h(r)$ could be set equal to $r^2$, without loss of generality. All other cases can be mapped to this case by an appropriate
coordinate transformation which is not the main topic of discussion here. The necessary condition for the existences of MSCOs
is given in terms of a second-order differential equation involving both $h(r)$ and $f(r)$ \cite{AMDP,TTN}
\begin{equation}
\frac{d}{dr}\Big (\frac{1}{f(r)}\Big ) \frac{d^2}{dr^2}\Big(\frac{1}{h(r)}\Big)
-\frac{d}{dr}\Big (\frac{1}{h(r)}\Big ) \frac{d^2}{dr^2}\Big(\frac{1}{f(r)}\Big)=0.
\end{equation}
For a given geometry of the black hole the functions $f(r)$ and $h(r)$ are specified and the above equation reduces to an algebraic equation in $r$, whose solutions provide us the required information about MCSOs. If in a spacetime, an MSCO exist then the solution of (2) gives us the radius of
such an orbit. The constants of motion $E$ and $L$ are given by \cite{TTN}
\begin{equation}
E^2=-\frac{1}{D}\frac{d}{dr}\Big (\frac{1}{h(r)}\Big ),
\end{equation}
\begin{equation}
L^2=-\frac{1}{D}\frac{d}{dr}\Big (\frac{1}{f(r)}\Big ),
\end{equation}
where
\begin{equation}
D=\frac{1}{h(r)}\frac{d}{dr}\Big (\frac{1}{f(r)}\Big )-\frac{1}{f(r)}\frac{d}{dr}\Big (\frac{1}{h(r)}\Big ) \neq 0.
\end{equation}
For any root of the equation (2) the sufficient condition for MSCO to exist, is that both constants of motion remains bounded, i.e.,
\begin{equation}
0\leq E^2 <\infty , \quad 0\leq L^2 <\infty. \label{ineq}
\end{equation}
If the condition given by (6) is satisfied for a root $r$ of (2) then
such a root is the radius of MSCO. Otherwise the root $r$ is unphysical. The existence of such orbits for both  Schwarzschild and Kottler black holes was proved in \cite{TTN} with the application of Strum's theorem. They also applied the analysis on spherically symmetric spacetimes in the Weyl conformal gravity. Here our purpose is to study the effect of quintessence on the stability of the circular orbits in a given black hole geometry. We now investigate the stability of such orbits in the Schwarzschild black hole geometry which is subjected to a quintessence field.

\section{MSCOs in qSBH}

For a Schwarzschild black hole surrounded by a quintessence matter field, the functions $f(r)$ and $h(r)$
in the equation (1) assumes the form \cite{VVK}
\begin{equation}
f(r)=1-\frac{r_g}{r}-\frac{\alpha}{r^{3\omega_q + 1}}, \quad h(r)=r^2.
\end{equation}
Here $r_g=2M$ and $M$ is the mass of the black hole and the normalization factor
satisfy $0<\alpha<1$ \cite{RNH}. Note that for $\alpha=0$, this black hole reduces
to the case of pure (without quintessence) Schwarzschild black hole. The critical power value $\omega_{q}=-1/3$,
reduces the quintessence term equal to a constant and we show that there exists an MSCO in this case. For
$\omega_q=-1$, this black hole spacetime becomes the Schwarzschild black hole
with the cosmological constant. For the horizon structure and other properties
of this black hole one my study for example \cite{SF,VVK}. Here we investigate
MSCOs in the Schwarzschild black hole with quintessence by studying the necessary
and sufficient conditions given by (2) and (6) for three different values of
$\omega_q$ appearing in $f(r)$ given by (7).

\subsection{qSBH with Equation of State ($\omega_q=-1/3$)}

The condition given in the equation (2) becomes
\begin{equation}
r-\alpha r-3r_g=0,
\end{equation}
where the case $\alpha=0$, gives the equation for the Schwarzschild black hole.
For unit mass of the black hole, $r_{g} =2$ and we get
\begin{equation}
r=\frac{6}{1-\alpha}\,. \label{msco1}
\end{equation}
In this case the constants of motion (3) and (4) simplifies into
\begin{align}
&E^2=-\frac{(2+r(\alpha -1))^2}{r(3+r(\alpha-1 )}\,,
\\
&L^2=-\frac{r^2}{3+r(\alpha-1)}\,,
\end{align}
Note that as we have assumed $r_{g}=2$, therefore in the above inequalities $r$ carries a dimensionless form. Using the equation (\ref{msco1}) we obtain
\begin{equation}
E^2=\frac{8}{9}(1-\alpha) \,.
\end{equation}
Since $\alpha <1$, thus the above quantity satisfy the first sufficient condition (6). We now investigate the second constant of motion which becomes
\begin{equation}
L^2=\frac{12}{(\alpha -1)^2 }\,.
\end{equation}
This is again a positive quantity. Therefore an MSCO exist for the qSBH at $r= 6/(1-\alpha)$. This will give the radius of an ISCO if $\alpha \rightarrow 0$. Similarly it specifies the radius of an OSCO if $\alpha \rightarrow 1$, which will be a sufficiently large number.

% \begin{equation}
% E^2=\frac{72(\alpha -1)}{8(1-\alpha)^3 -108}\,.
% \end{equation}
% Thus the root given by equation (8) is physical if the sufficient condition is satisfied
% \begin{equation}
% \frac{72(\alpha -1)}{8(1-\alpha)^3 -108}\geq 0.
% \end{equation}
% Since $0<\alpha<1$, hence $\alpha -1 <0$, therefore the above inequality holds only if the denominator is negative, i.e.,
% \begin{equation}
% 8(1-\alpha)^3 -108 < 0.
% \end{equation}
% This implies that
% \begin{equation}
% 8(1-\alpha)^3 -108 < 0.
% \end{equation}
% This leads to $\alpha\geq 1$. This is a contradiction and hence there is no MSCO in this case.

\subsection{qSBH with $\omega_q =-1$ (Cosmological Constant) }

In this case the condition (2) takes the following form
\begin{equation}
8\alpha r^4-15\alpha r_g r^3-r_g r+3 r_g^2=0.
\end{equation}
It is convenient to write above equation in a dimensionless form, we introduce
$x=r/r_g$ and $\lambda=\alpha r_g^2$. Then (14) becomes
\begin{equation}
8\lambda x^4-15\lambda x^3-x +3=0.
\end{equation}
The above equation coincides with the equation obtained for the Schwarzschild-de
Sitter spacetime \cite{TTN}. From the applications of Sturm's theorem Toshika et. al
\cite{TTN}, have shown that only for $0<\lambda<16/16875$, the necessary and sufficient
conditions (2) and (6) are satisfied and two MSCOs exist. There is one ISCO
and the other is the OSCO. From this we obtain bounds on the normalization factor
as $0<\alpha<4/16875$. Thus we can say that when $\omega _q=-1$, then for the unit mass
of the the Schwarzschild black hole with quintessence the MSCOs exist if $0<\alpha<4/16875$.

\subsection{qSBH as a Kiselev Black Hole ($\omega_q=-2/3$)}

This is a simplest nontrivial case of the Kiselev black hole \cite{VVK}, where
the geometry is governed by the line element
\begin{equation}
 \mbox{d}s^2=-\Big ( 1- \frac{r_{g}}{r}-\alpha r \Big )\mbox{d}t^2+\frac{\mbox{d}r^2}{\Big ( 1- \frac{r_{g}}{r}-\alpha r \Big )}+r^2(\mbox{d}{\theta}^2+\sin^2{\theta}\mbox{d}\phi^2 )\,,
\end{equation}
which has the scalar curvature $R= 6\alpha/r$, where the inner and outer horizons exist at the coordinate singularities
\begin{align}
&r_{in} = \frac{1 - \sqrt{ 1- 4 \alpha r_{g} }}{2 \alpha} , \quad 	r_{out} =\frac{1 + \sqrt{ 1- 4 \alpha r_{g}}}{2 \alpha} \,,
\end{align}
provided $4r_g \alpha < 1$ and the two coincides if $4r_g \alpha = 1$, in which case we arrive at an extremal black hole. For a unit mass black hole, it results into $\alpha < 1/8$ and $\alpha =1 /8$, respectively.

The problem is to find the positive roots ($r>0$) of the equation (2) with $f(r)$ and $h(r)$, defined in the above metric such that it also satisfies two additional constraints (6) in the form of inequalities given by
\begin{align}
&
E^2 = -\frac{2(\alpha r^2 -r +r_{g})^2}{r(\alpha r^2 -2r+3r_{g})} >0\,, \label{eqa}\\
&
L^2 = -\frac{r^2\Big (r_{g}-\alpha r^2 \Big)}{\alpha r^2 -2r+3r_{g}} >0\,. \label{eq1}
\end{align}
Before solving equation (2) for the above Kiselev black hole it is better to obtain the critical bounds on $r$ on the positive $r-$axis for which there exists a solution. Let us determine the critical bounds on $r$ using sufficient conditions. Note that in the first inequality ($\ref{eqa}$) the numerator is a positive number so the inequality is true only if
\[
\alpha r^2 -2r+3r_{g} < 0, \label {ineq0}
\]
which can be factorized as follows
\[
(r-r_{-})(r-r_{+}) < 0 ,\label {ineq1}
\]
where $r_{-}$ and $r_{+}$ are given below
\[
r_{-} = \frac{1- \sqrt{1-3\alpha r_{g}}}{\alpha} \,, \quad r_{+} = \frac{1+ \sqrt{1-3\alpha r_{g}}}{\alpha} \,.
\]
This implies that $\alpha <1/3r_{g}$, as otherwise we obtain complex constants of motion and both inequalities (\ref{eqa}) and (\ref{eq1}) are violated. In order to hold true the inequality (\ref{ineq0}) implies that both factors in (\ref{ineq1}) have opposite signs which is only possible when the value of $r$ lies in the interval $\big( r_{-}, r_{+} \big)$. Since $\alpha$ is arbitrary therefore it can be used to identify the local bounds on $r_{-}$ and $r_{+}$, for which (\ref{ineq0}) is valid. We now employ Taylor expansion as $3r_{g}\alpha <1$, to get
\begin{align}
r_{-} &= \frac{1}{\alpha} \left ( 1- \left ( 1-\frac{3\alpha r_{g}}{2} - \frac{(3\alpha r_{g})^2}{8}-  \mbox{O}(\epsilon^3) \right ) \right), \\
& = \frac{3r_{g}}{2}+  \frac{9r_{g}^2 \alpha}{8} +\mbox{O}(\epsilon^3), \quad \epsilon = 3\alpha r_{g} .
\end{align}
Therefore the least value of $r_{-}$ is $3r_{g}/2$, as $ r_{-} >3r_{g}/2$. Similarly
\begin{align}
r_{+} &= \frac{1}{\alpha} \left ( 1 + \left ( 1-\frac{3\alpha r_{g}}{2} - \frac{(3\alpha r_{g})^2}{8}-  \mbox{O}(\epsilon^3) \right ) \right), \\
& = \frac{2}{\alpha} - \frac{3r_{g}}{2} -  \frac{9r_{g}^2 \alpha}{8} +\mbox{O}(\epsilon^3),
\end{align}
thus $r_{+}<2/\alpha$, so the largest value of $r_{+}$ is $2/\alpha$, hence we obtain
\[
r \in \Bigg( \frac{3r_{g}}{2}+  \frac{9r_{g}^2 \alpha}{8} \,,\, \frac{2}{\alpha} \Bigg) \,,
\]
where the first constant of motion satisfy $E^2>0$, in (18). Since the denominator in the first inequality (18) is negative therefore the other inequality (19) holds true if we have
\[
r_{g}-\alpha r^{2} > 0,
\]
which holds when
\[
 \sqrt{\frac{r_{g}}{\alpha}} ~>~ r\,.
\]
Since $r_{+} < 2/\alpha$, which follows from the condition (27) and as
\[
\sqrt{\frac{r_{g}}{\alpha}}~< ~ \frac{2}{\alpha} ~ \,,~\forall~~\alpha <\frac{1}{4r_{g}}
\]
therefore the upper bound $r_{+}$ is irrelevant for our purpose. Thus both inequalities (\ref{eqa}) and (\ref{eq1}) hold true provided the following lemma holds.

\textbf{Lemma 1.} \emph{For a unit mass Schwarzchild black hole ($M=1$) surrounded by a quintessence field with inner and outer horizons, the roots of equation (2) satisfy the inequalities (\ref{eq1}), provided they satisfy the global bound
\[
r \in \Bigg(\frac{3r_{g}}{2}+  \frac{9r_{g}^2 \alpha}{8}\,,\,\sqrt{\frac{r_{g}}{\alpha}} \Bigg) \,. \label{lem}
\]
}
The above lemma provides a range on $r$ for which both constants of motion satisfy the positivity criteria (18) and (19). Note that for an extremal Schwarzschild black hole surrounded by quintessence the above range reduces to
\[
r \in \Bigg( \frac{3r_{g}}{2} +  \frac{9r_{g}^2 }{64} \,,\, 4   \Bigg) ,
\]
which for a unit mass of the black hole gives $r\in( 3.5625 \,,\, 4)$.

Note that since $\alpha< 1/3r_{g}$, therefore the upper bound in (\ref{lem}) sharply moves away from origin as the value of $\alpha$ is chosen close to zero. However the lower bound contains $\alpha$ in the numerator which only contribute a small change in it. Therefore we expect to get more number of MSCOs in our analysis if the value of $\alpha$ lies in the close vicinity of zero. Indeed further analysis provide us sharper bounds on the values of $\alpha$ that yields a complete classification of MSCOs in Kieslev black holes. Thus, we only look for the roots of algebraic equation which satisfy Lemma 1. In this case, we obtain the following constraint from equation (2)
\begin{equation}
\alpha^2 r^4-3\alpha r^3+6\alpha r_g r^2+ r_g r -3r_g^2=0.
\end{equation}
As required, for $\alpha=0$, the above equation reduces to the equation for the Schwarzschild balck
hole. We write it in dimensionless form by defining $x=r/r_g$ and $\lambda=\alpha r_g$
\begin{equation}
\lambda^2 x^4-3\lambda x^3+6\lambda x^2+x -3=0. \label{eq0}
\end{equation}
We now proceed to find the roots of quartic equation (\ref{eq0}) using Maple and it turns out that these can be converted into radicals
\begin{equation} \label{rad}
x= \frac{3 \sigma_{\omega}^{1/6} \kappa_{\sigma}^{1/4} + \kappa_{\sigma}^{3/4} + \sqrt{ \vphantom{sum}(54-160 \lambda)\sigma_{\omega}^{1/2}+ (( 18-32\lambda ) \sigma_{\omega}^{1/3}-4\lambda^{1/2})\kappa_{\sigma}^{1/2}
- 4 (\lambda \kappa_{\sigma})^{1/2}\sigma^{2/3}  }}{4 \kappa_{\sigma}^{1/6} \sigma_{\omega}^{1/4}} \,,
\end{equation}
where $\omega$, $\kappa_{\sigma}$ and $\sigma_{\omega}$ are defined below
\begin{align} \label{ceq1}
& \omega = {1024\lambda^3 -640\lambda^2+100\lambda-1 } \,,\\
&\sigma_{\omega} = 32\lambda^{3/2}+\sqrt{\omega} -10\sqrt{\lambda}\,, \label{ceq2}\\
&\kappa_{\sigma} = 4\sqrt{\lambda}\,\sigma_{\omega}^{2/3}-16\sigma_{\omega}^{1/3}\lambda +9\sigma_{\omega}^{1/3}+4 \sqrt{\lambda}\,\label{ceq3}.
\end{align}
Note that the main term in the radical (\ref{rad}) is $\omega$, while rest of the terms $\sigma_{\omega}$ and $\kappa_{\sigma}$ are defined in terms of it. Furthermore, the main equation of $\omega$, depends entirely on $\lambda$, which can be used to characterize the ranges of $\lambda$ that yield feasible MSCOs. Therefore, we now consider $\omega$, as a function of $\lambda$,
\begin{equation}
\omega (\lambda) = 1024\lambda^3 -640\lambda^2+100\lambda-1 ,
\end{equation}
which is a third degree equation in $\lambda$. We now examine the behavior of function $\omega(\lambda)$ in terms of $\lambda$, where we already have that $\lambda >0$. Note that $\omega(\lambda)$ starts with a negative value $\omega(0) =-1$, and increases afterwards therefore the exact value of $\lambda$ where after the graph of $\omega(\lambda)$ is positive can be obtained by finding the roots of above equation, i.e. $\omega(\lambda)=0$ which gives
\begin{equation}
\lambda_{1} = \frac{3-2\sqrt{2}}{16},~\lambda_{2}=\frac{1}{4}, ~\lambda_{3}=\frac{3+2\sqrt{2}}{16}\,.
\end{equation}
Note that the above values of $\lambda$ satisfy $\lambda_{1}<\lambda_{2}<\lambda_{3}$, therefore the least value of $\lambda$ is $\lambda=\lambda_{1}= 0.0107233048$. The graph of the function is given in Figure 1. From the graph it is clear that $\omega (\lambda)>0$ for all $\lambda\in (\lambda_{1},\lambda_{2})$, after which it again becomes negative in the interval $\lambda\in (\lambda_{2},\lambda_{3})$. Lastly for all $\lambda \in (\lambda_{3}, 1)$, the function $\omega$ is positive. We have found that in the intervals where $\omega$, is positive gives rise to two real (in which one is positive and the other is negative) and two imaginary roots. However, there are three positive and one negative real root in the ranges where $\omega$ is negative. In short the feasible regions include a union of two disjoint intervals, i.e.
\begin{equation}
\omega(\lambda) > 0 , ~ \forall \quad \lambda \in \big( 0, 0.0107233048 \big ) \cup \big(0.25,0.3642766952  \big) ,
\end{equation}
which in terms of the value $\alpha$ becomes (where $\lambda = 2\alpha$)
\begin{equation}
\omega(\alpha) > 0 , ~ \forall \quad \alpha \in \big (0, 0.00536165238 \big) \cup \big(0.125, 0.1821383476 \big).
\end{equation}
\begin{figure}[h]
\centering
\includegraphics[width=2.8in]{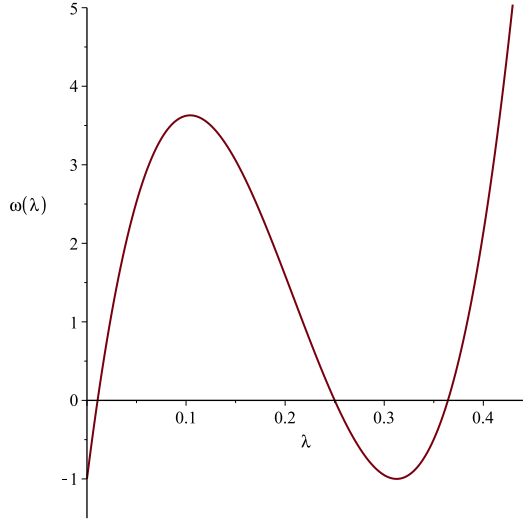}
\caption{The graph of $\omega(\lambda)$, where it has three real roots. }
\end{figure}
We now examine above regions in the light of Lemma 1. Note that $\alpha<1/3r_{g}$, which for a unit qSBH becomes $\alpha< 1/6$, i.e. $\alpha < 0.1666666667.$ Since the value $0.1666666667$ is smaller than $0.1821383476$, therefore it provides a sharper upper bound on the range of $\alpha$. It is easy to verify that in the interval $\alpha \in \big(0.125, 0.1666666667 \big)$, we obtain four MSCOs (three positive and one negative) all of which fail to satisfy Lemma 1. Therefore, the only interval of interest is $\big (0, 0.00536165238 \big)$, in which the condition of Lemma 1 is also fulfilled. In order to find the MSCOs we have to solve the quartic equation (\ref{eq0}) and we find that in the regions where $\omega(\lambda)$ is positive it yields two real and two imaginary roots. On the other hand in the regions where $\omega(\lambda)$ is negative the equation results into three positive and one negative real root. We now consider a few particular cases to show that MSCOs exist. For example assume that $\lambda=0.002$, i.e., $\alpha=0.001$,
then (\ref{eq0}) becomes
\begin{equation}
0.000004 x^4-0.006 x^3+0.012 x^2+x -3=0.
\end{equation}
The real positive roots of this equation are obtained after using (\ref{ceq1}-\ref{ceq3}) are
\begin{equation}
x_1=3.059118379, \quad x_2=12.32978727, \quad x_{3} = 1497.885976,
\end{equation}
where we have discarded one negative root. Now for unit mass of the black hole this gives
\begin{equation}
r_1=6.118236758, \quad r_2=24.65957454, \quad r_{3} = 2995.771952,
\end{equation}
which means that both $r_{1}$ and $r_{2}$ lies in the required interval $(3, 44.72135)$ of Lemma 1. Since, $r_{3}$ does not lie in the interval therefore fail to satisfy Lemma 1. Therefore here we have two MSCOs. One as ISCO with $r_{ISCO}=6.119$ and the other as OSCO with $r_{OSCO}=24.66$.

For another value of $\lambda=0.01$, i.e. for $\alpha=0.005$, we get two
MSCOs. The ISCO with $r_{ISCO}=7.2378$ and OSCO with $r_{OSCO}=9.1628$.
One can easily check that for other values $\alpha=0.002$, $\alpha=0.003$
and $\alpha=0.004$ there exist MSCOs. Similarly if we consider $\alpha=0.006$, then we get one positive
real root of (2) for which
$E^2<0$. Hence it is unphysical and no MSCO exist.

\section{Summary}

In this brief communication we have analyzed the MSCOs of a test
particle in the vicinity of the Schwarzschild black hole with quintessence
matter to obtain bounds on the value of the normalization factor $\alpha$,
for which such orbits exist. We have considered three different cases for
the value of the equation of state parameter, i.e. $\omega_q=-1$, $\omega_q=-1/3$
and $\omega_q=-2/3$. In the case when $\omega_q=-1$, we have
obtained $0<\alpha<4/16875$, for which there exist two MSCOs. For $\omega_q=-1/3$,
we have seen that MSCO exists for all $\alpha \in (0,1)$. While in the case of $\omega_q=-2/3$, we see
that MSCOs exist if $0<\alpha\leq 0.00536165238$.

Another observation is that in the presence of a quintessence field ($w_{q}=-1,-2/3$), the radius of
the MSCOs gets lager as compared to the radius of MSCO of a pure unit mass Schwarzschild black hole
for which it is $6$. Recently the effect of a quintessence model on the energy content of the Riessner-Nordstrom
black hole surrounded by the quintessence matter has been investigated in \cite{IS}.
Both upper and lower bounds were obtained on the value of the normalization factor
$\alpha$, i.e. $0<\alpha<1$. The same bonds on the normalization factor $\alpha$
were also obtained for the Schwarzschild black hole with quintessence matter in a
totally different scenario \cite{RNH}. Therefore, it is worth exploring to check
whether the same bounds on the normalization factor $\alpha$ obtained here, also exist in the case of the Riessner-Nordstrom black hole with quintessence matter.

\section*{Acknowledgments} IH is very grateful to Kavli Institute for Theoretical Physics,
Chinese Academy of Sciences, Beijing, China, where this work was initiated under the TWAS-UNESCO
Associateship.

\end{document}